\documentclass[a4paper]{jpconf}
\usepackage{graphicx}
\begin{document}

\title{High-sensitivity tool for studying phonon related mechanical losses in low loss materials}

\author{Daniel Heinert$^1$, Anja Zimmer$^1$, Ronny Nawrodt$^1$, Torsten Koettig$^1$, Christian Schwarz$^1$, Matthias Hudl$^1$, Wolfgang Vodel$^1$, Andreas T\"unnermann$^2$ and Paul Seidel$^1$}

\address{$^1$ Institute of Solid State Physics, Friedrich-Schiller-University Jena, Helmholtzweg~5, D-07743~Jena, Germany}
\address{$^2$ Institute of Applied Physics, Friedrich-Schiller-University Jena, Max-Wien-Platz~1, D-07743~Jena, Germany}

\ead{anja.zimmer@uni-jena.de}

\begin{abstract}
Fundamental mechanical loss mechanisms exist even in very pure materials, for instance, due to the interactions of excited acoustic waves with thermal phonons. A reduction of these losses in a certain frequency range is desired in high precision instruments like gravitational wave detectors. Systematic analyses of the mechanical losses in those low loss materials are essential for this aim, performed in a highly sensitive experimental set-up. Our novel method of mechanical spectroscopy, cryogenic resonant acoustic spectroscopy of bulk materials (CRA spectroscopy), is well suited to systematically determine losses at the resonant frequencies of the samples of less than $10^{-9}$ in the wide temperature range from 5 to 300 K. A high precision set-up in a specially built cryostat allows contactless excitation and readout of the oscillations of the sample. The experimental set-up and measuring procedure are described. Limitations to our experiment due to external loss mechanisms are analysed. The influence of the suspension system as well as the sample preparation is explained.
\end{abstract}

Especially materials offering low mechanical losses require highly sensitive mechanical spectroscopy methods for studying their internal loss processes. Such materials are for instance used in high precision instruments for the detection of gravitational waves \cite{Sau1994}. Hypothetically a reduction of mechanical losses can be achieved by producing a total defect-free material. But even if switching over from a real to an ideal solid, losses still remain. They are caused by interactions of excited acoustic waves with thermal phonons and eventually free electrons (besides thermoelastic losses). Therefore minimizing mechanical losses in the mentioned applications requires studying these fundamental loss mechanisms. Intrinsic loss processes in pure materials are best examined by using bulk materials. A novel method of mechanical spectrocopy on bulk materials, cryogenic resonant acoustic spectroscopy of bulk materials (CRA spectroscopy), has been successfully tested on the fairly well-known material crystalline quartz \cite{Zim2007}. A further characterization of the method with focus on the suspension system is presented in this paper. In CRA spectroscopy the reciprocals of the mechanical losses at the resonant frequencies are measured, the mechanical quality factors Q. One aim of the method is to extrapolate losses at other frequencies by performing systematic Q measurements.

\section{Experimental set-up and measurement technique}

The measurements of the mechanical Q factors are performed within a temperature range from 6~K to 300~K. The probe chamber is located in a specially built cryostat \cite{Naw2006}. In combination of cooling with the boil-off gas from the liquid helium tank and heating the desired temperature can be achieved by means of a LakeShore temperature controller (LS340) and is kept during the Q measurements to better than 0.3~K. A non-contact measurement of the temperature of the first sample is done by determining the temperature of a 'twin substrate' which is symmetrically placed to the first one and has a calibrated temperature sensor attached to it.
To reduce residual gas damping the probe chamber is evacuated to at least 10$^{-3}$ Pa. The sealing is done by metallic indium.\\
The suspension of the substrate under investigation should interact as less as possible with the sample. A transfer of energy from the sample to the suspension would result in additional damping. Different suspension systems have been proposed. The suspension of the substrate by means of a wire or ribbon as a pendulum \cite{Gil1993}, by clamping it between two nodal points \cite{Num2000} or even balancing it on a hemisphere under certain conditions \cite{Gep}. We focused on the pendulum suspension. For details see section \ref{susp}.\\
In this case, the Q factor is determined by an amplitude ring-down method. First, the substrate is excited at one of its resonances.
For higher mechanical losses ($>10^{-4}$) the Q factor is determined out of the 3dB-width of the resonance peak. We mainly deal with low mechanical losses where the half-width of the resonance peak is very small.
This is possible without any contact by using comb-like electrodes at a high voltage (up to 1600~V) with the corresponding frequency. The magnitude of the oscillation amplitude depends on the dielectric properties of the substrate material. For crystalline quartz, silicon, CaF$_{2}$ and fused silica amplitudes of 0.2~nm to 1000~nm are achieved. Then, the driving electric field is removed and the subsequent amplitude ring-down is
recorded. The readout of the decaying amplitude is also arranged to be contactless. 
\begin{figure}[htb]
a)~\includegraphics[width=.5\textwidth]{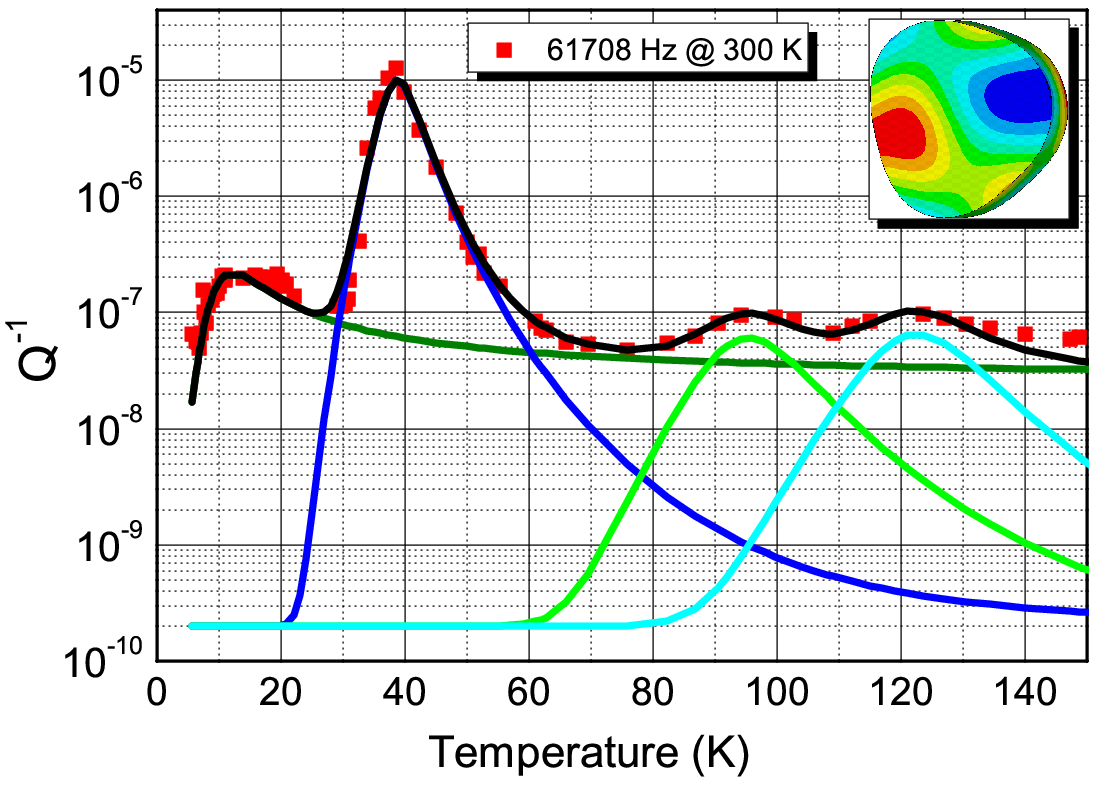}
\hspace{1cm}
b)~\includegraphics[width=.21\textwidth]{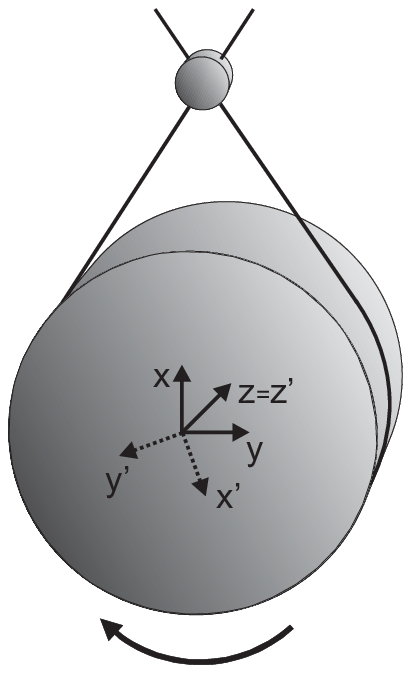}
\caption{a) Damping versus temperature gained by Q measurements on crystalline quartz (z-cut) and corresponding fits of damping contributions. The excited mode shape is shown as a contour plot of the displacement in direction of the cylinder axis. Red and blue denote maximum displacement, green minimum. Red squares: reciprocals of measured Q factors, dark green line: contribution to damping by thermal phonons; dark blue, light green and light blue lines: contribution due to defect induced hopping. For further information, including fit parameters, see the work of Zimmer et al. \cite{Zim2007}. b) Schematic view of the suspension system. Samples of anisotropic materials might have to be rotated in the wire loop to gain the highest Q factor possible in the suspension system.}
\label{fig:1}
\end{figure}
A window in the cryostat permits the usage of a  Michelson-like interferometer which has a resolution of about 0.1~nm up to a frequency of 500~kHz.
The recorded 1/e-ring-down time is  converted into the Q factor. Repeated measurements have shown that the Q factor is determined with an accuracy of 10~\% to 20~\%.\\
If the liquid helium holding time of the cryostat is spend for measuring a total 1/e-ring-down time, this 30 hours correspond to a Q factor of $Q\approx3.3\times10^{9}$ taken for the modes with the lowest frequencies of about 10~kHz. Clear signals reduce the required recording time to ring-down to 90~\% of the initial amplitude. The highest potentially measurable Q factor (assuming a 10~kHz mode) is therefore increased by factor 10 to $Q\approx3.3\times10^{10}$.\\
For an example of the dominating contribution of the interactions of the acoustic wave with thermal phonons to the damping over the whole temperature range see a measurement on a crystalline quartz sample (cylinder, 74.8~mm in diameter and 12.05~mm thick) in Fig. \ref{fig:1} a).

\section{Sample preparation}

The samples are cut to the desired orientation with an accuracy of better than one degree. This work as well as polishing of the samples was done by the company Hellma-Optik \cite{Hellma}, which is specialized on polishing of optical components. As a result of this process the crystal exhibits a roughness lower than 10~nm on all surfaces. After this geometrical manufacturing, the sample passes through a standard optical cleaning process before each measurement. In a clean room facility it is treated with acetone and isopropanol.

\section{Influence of the suspension system}\label{susp}

The substrate is, under the usage of rubber gloves, carefully put into the suspension consisting of a tungsten wire loop. For this purpose the tungsten wire is polished with diamond paste and afterwards cleaned with isopropanol. Experiments show, that the location of the wire on the circumference is crucial for the measurement of high Q factors. Especially, a position that symmetrically divides the crystal into halves seems advantageous. A further influence on the measured Q factor exists in tilting of the substrate in the wire loop. A tilting of few arc minutes may reduce the Q factor.\\
\begin{figure}[htb]
\centering
\includegraphics[width=.7\textwidth]{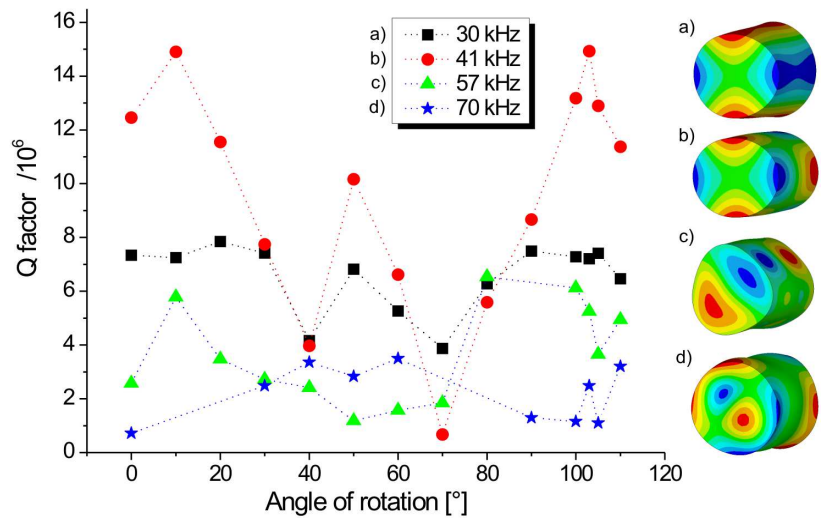}
\caption{Q factor versus angle of rotation of the substrate in the wire loop. The dotted lines are drawn as guide line for the eye. The corresponding mode shapes are plotted on the right. A contour plot showes the displacements in direction of the cylinder axis. Red and blue areas indicate maximum displacement whereas green areas are motionless.}
\label{fig:rot}
\end{figure}
The wire's diameter is chosen to be preferably thin, such that the stress under load of the substrate is higher than 50\% of the rupture stress. This choice originates in the relationship between stress and dissipated energy in the suspension wire. The dilution factor expresses this behavior quantitatively. For further information see, e.g, the work of Cagnoli et al. \cite{Cag2000}. Close above the substrate the wire is fixed with an aluminium clamp in a crossed position. To provide a stable configuration wich is essential for cryogenic measurements the crossing point is located in the middle of the clamp. The length of the wire should be chosen in a way that resonances of the suspension do not coincide with excited resonances of the substrate. For further information see the work of Nawrodt et al. \cite{Naw2007}.\\
The optimization of the suspension system follows with regard to the determined decay time and quality factor as the last step. For isotropic materials this step can be skipped as rotations of the substrate in the wire loop leave the Q factors unchanged. The mode shape is self-aligned with respect to boundary conditions minimizing friction loss . In contrast, in anisotropic materials the mode shape rotates with the structure coordinate system. Depending on the coupling of the displacements of the substrate to the wire the Q factor varies with orientation of the substrate in the wire loop. To find the highest possible Q factor of the substrate in this suspension system the substrate is rotated by small angles (Fig. \ref{fig:1} b)). The position corresponding to the highest Q factor is that where the least energy is transferred from the substrate to the suspension. For results on the variation of the Q factor with angle of rotation see Fig. \ref{fig:rot}. The starting position was arbitrarily chosen. The readout of all Q measurements belonging to the same rotation angle was done on the same point on the surface. The substrate under investigation was made of CaF$_{2}$ (100) with a diameter of about 75~mm and a thickness of also 75~mm. Q measurements versus temperature on the same sample are shown in the work of Zimmer et al. \cite{Zim2007b}. Although the angle was roughly estimated with up to 10$^\circ$ error, a symmetry around 50$^\circ$ is clearly visible for all modes. This symmetry is reflected in the mode shapes. Thus, about 30 runs of rotation are necessary to start the Q measurements with the highest achievable Q factor in the suspension system. Nevertheless, the benefit of deeper knowledge about thermal phonons und defects in low loss materials justifies the mentioned challenges of experimental implementation.

\ack This work was supported by the German DFG under contract SFB Transregio 7.

\end{document}